\documentclass[a4paper,fleqn,usenatbib]{mnras}

\usepackage[T1]{fontenc}
\usepackage{ae,aecompl}

\usepackage{graphicx}	
\usepackage{amsmath}	
\usepackage{amssymb}	
\newcommand{\nicer}{\textit{NICER}~}

\newcommand{\swift}{\textit{Swift}~}
\newcommand{\eps}{erg s$^{-1}$~}

\title[BH Accretion]{Global Accretion Properties of Black Hole X-Ray Binaries: A Phenomenological Perspective}

\author[Jana et al.]{
Arghajit Jana$^{1}$\thanks{E-mail: janaarghajit@mx.nthu.edu.tw, argha0004@gmail.com},
\\
$^{1}$Institute of Astronomy, National Tsing Hua University, Hsinhcu, 30013, Taiwan\\
}
\date{Accepted XXX. Received YYY; in original form ZZZ}

\pubyear{2020}

\begin{document}

\label{firstpage}
\pagerange{\pageref{firstpage}--\pageref{lastpage}}
\maketitle

\begin{abstract}
Black hole X-ray binaries (BHXBs) show rich phenomenology in the spectral and timing properties. We collected the spectral data of 20 BHXBs from the literature across different spectral states. The spectral properties are studied in the forms of the inner disc temperature ($T_{\rm in}$), photon index ($\Gamma$), hot electron temperature ($kT_{\rm e}$), X-ray flux ($F_{\rm X}$) and luminosity ($L_{\rm X}$). We studied various correlations among different spectral parameters to understand the accretion process on a global scale. In the thermal soft states (TSS), we find most of the sources followed $F_{\rm disc} \propto T_{\rm in}^4$ relation. A `V'-shaped correlation is found between $\Gamma$ and total luminosity ($L_{\rm tot}$) in the hard Comptonized state (HCS). The Comptonized luminosity is observed to be correlated with the disc luminosity in the HCS and TSS. No notable correlation is observed in the intermediate state (IMS). The evolution of the inner disc radius ($R_{\rm in}$) is unclear in the HCS and IMS. We also discuss how the hot electron temperature changes with other spectral parameters. We observe that the iron line flux correlates with disc and Comptonized fluxes. The strength of the reprocessed emission is found to vary across spectral states.

\end{abstract}



\section{Introduction}
\label{sec:intro}
A transient black hole X-ray binary (BHXB) spends most of the time in a quiescent state with very low X-ray luminosity ($L_{\rm X} < 10^{32}$ \eps) and occasionally goes to outburst that lasts from several weeks to months \citep[e.g.,][]{RM06,Tetarenko2016}. The X-ray luminosity increases a few orders of magnitude during the outburst compared to the quiescent state. On the other hand, a persistent source is found to be always active with the X-ray luminosity $L_{\rm X} > 10^{35}$ \eps. 

An X-ray spectrum of BHXBs can be approximated by a multi-colour blackbody (MCD) component and a power-law tail (PL). The geometrically thin and optically thick accretion disc is believed to be the origin of the MCD \citep{SS73}, while the hard PL tail originates in a Compton corona, located close to the BH \citep[e.g.,][]{HM93,CT95,Done2007}. A fraction of the seed photons up-scatter with the hot electrons of the Compton corona and produce a power-law tail via inverse-Comptonization \citep{ST80,ST85}. Further, a fraction of the hard photons is reprocessed in the disc and produces an iron K$\alpha$ line at $\sim 6.4$~keV and a reflection hump at $\sim 20-40$~keV \citep[e.g.,][]{Fabian1989,Matt1991}.

The BHXBs also exhibit low-frequency quasi-periodic oscillations (LFQPOs) in the power-density spectra (PDS) observed in a range of $0.1-20$~Hz \citep[see, ][for a review]{RM06,Ingram2019}. Depending on the the Q-factor($Q=\nu/\Delta \nu$, $\nu$ and $\Delta \nu$ are centroid frequency and full width at half-maximum; FWHM), the rms amplitude, a LFQPO can be classified as Type-A, Type-B, or Type-C \citep[][and references therein]{Casella2005}.

During the outburst phase, a BHXB shows fast variability and fluctuation in the spectral and timing properties \citep[e.g.,][]{vanderklis1989,vanderklis1994,Mendez1997}. A correlation between the spectral and timing properties is seen in the hardness–intensity diagram \citep[HID; e.g.,][]{Homan2001,Homan2005}, accretion rate–intensity diagram \citep[ARRID; e.g.,][]{Mondal2014,AJ2016}, rms-intensity diagram \citep[RID; e.g.,][]{Munoz-Darias2011} or hardness–rms diagram \citep[HRD; e.g.,][]{Belloni2005}. Commonly, two major spectral states are observed in BHXBs: hard Comptonized state (HCS) and thermal soft state \citep[TSS; e.g.,][]{Belloni2005,RM06,Nandi2012}. In between HCS and TSS, an intermediate state (IMS) is also observed \citep[e.g.,][]{RM06}. A BHXB transits from the HCS to TSS or vice-versa through the IMS, i.e., IMS acts as the state transition phase. Sometimes, the IMS is further divided into hard intermediate state (HIMS) and soft-intermediate state (SIMS). An outbursting BHXB is generally found to evolve as HCS $\xrightarrow{}$ HIMS $\xrightarrow{}$ SIMS $\xrightarrow{}$ TSS $\xrightarrow{}$ SIMS $\xrightarrow{}$ HIMS $\xrightarrow{}$ HCS. A BHXB goes through all the spectral states in a complete outburst. However, some outbursts do not show TSS and are known as the `failed' outburst \citep[e.g.,][]{Tetarenko2016,DD2017}.

A BHXB shows rich phenomenology across all the spectral states. Each spectral state is characterized by different spectral and timing properties \citep[e.g.,][]{RM06,Done2007}. A HCS spectrum is characterized by a cool disc of temperature $T_{\rm in} \sim 0.1-0.5$~keV, and photon index, $\Gamma \sim 1.5-1.7$. The disc component is not observed sometimes in the HCS. The hard X-ray photon flux dominates over the soft photon flux in this state. BHXBs also exhibit a compact and quasi-stable jet in the HCS \citep[e.g.,][]{Fender2004}. In this state, an evolving type-C QPO is observed. The TSS spectra can be described by a disc of a temperature, $T_{\rm in}\sim 1$~keV and photon index, $\Gamma > 2.3$. No jet is observed in this state. QPOs also are not observed in the TSS. The source transit to the TSS from the HCS via an IMS, i.e. HIMS and SIMS. In these spectral states, the soft and hard photon fluxes are comparable. The IMS is often seen to be associated with the discrete ejection or blobby jet \citep[e.g.,][]{Nandi2001,Fender2004}. Evolving type-C QPOs are generally observed in the HIMS, while sporadic type-A or type-B QPOs are observed in the SIMS.

The phenomenology of the BHXBs can be studied in terms of the disc temperature ($T_{\rm in}$), photon index ($\Gamma$) and X-ray flux ($F_{\rm X}$). Correlations among different spectral parameters and the evolution of the timing and spectral parameters are well established. Here, we revisited the phenomenological properties and different correlations among them to test if these relations hold globally. Here, we gathered information on the spectral parameters for 20 BHXBs across different spectral states. We studied various correlations among the sources individually and with the entire sample across all the spectral states. The paper is organized in the following way: In \S2, we described the sample. Then, the analysis process and results are presented in \S3. Finally, in \S4, we discussed our findings.

\begin{table*}
\centering
\caption{List of Sources}
\begin{tabular}{ccccccccccc}
\hline
No & Name & Mass  &  Dist & Incl.  & Outburst  & Telescope & Energy Range & Ref. \\
 & & ($M_{\odot}$) & (kpc) & (deg) & & & (keV) & \\
\hline
1 & GRS 1716--249      &$  5.0\pm0.7      $& $ 2.4\pm0.4 $& $40-50$& 2017 & Swift/XRT & $0.5-10$ &  1,2,3, 4  \\
  &                    &                   &              &        &      & Swift/BAT & $15-185$ & \\
2 & GX 339--4          &$  9.0\pm1.5      $& $ 8.4\pm0.9 $& $30\pm1$& 1996-1999, 2002, 2004 & RXTE/PCA & $2-25$ & 5, 6, 7, 8, 9 \\
  &                    &                   &               &        & 2007, 2010-2011 & RXTE/HEXTE & $18-200$ & \\        
3 & H 1743--322        &$  11.2\pm1.9     $& $ 8.5\pm0.8 $& $75\pm3$& 2003, 2008, 2009, & RXTE/PCA & $2-25$ & 10, 11, 12, 13,\\
  &                    &                   &              &         & 2010a, 2010b, 2011 & RXTE/HEXTE & $24-200$ & 14, 15, 16\\
4 & MAXI J0637--430    &$  8.5\pm3.5      $& $ 8.0       $& $64\pm6$& 2019-20 & NICER & $0.7-10$ &  17, 18 \\
5 & MAXI J1348--630    &$ 9.1\pm1.6       $& $2.2 \pm 0.6$& $29.2\pm0.5$ &2019 & Swift/XRT & $0.5-10$ & 19, 20, 21  \\
  &                    &                   &              &        &      & Swift/BAT & $15-150$ & \\
6 & MAXI J1535--571    &$ 8.9\pm1.0       $& $4.1\pm0.5  $& $57\pm2$ & 2017 & Swift/XRT & $0.5-10$ & 22, 23, 24, 25  \\
7 & MAXI J1659--152    &$  6.0\pm1.5      $& $ 8.6\pm3.7 $& $70-80$ &2010 & RXTE/PCA & $2-25$ & 26, 27, 28, 29 \\
8 & MAXI J1727--203    &$  10\pm2         $& $5^*$& $45^*$ & 2018 & NICER & $0.5-10$ & 30 \\
9 & MAXI J1813--095    &$  7.4\pm1.5      $& $ 6.0       $& $28-45$ & 2018 & NICER & $0.5-10$ & 31, 32\\
  &                    &                   &              &        &      & Swift/XRT & $0.5-10$ & \\
  &                    &                   &              &        &      & NuSTAR & $3-78$ & \\
10 & MAXI J1836+194    &$  9.5\pm1.5      $& $ 7.0\pm3.0 $& $4-15$ & 2011 & RXTE/PCA & $2-25$ & 33, 34, 35  \\
11& Swift J1357.2--0933&$ 5.4\pm1.4       $& $1.5        $& $70$ & 2011 & Swift/XRT & $0.5-10$ & 36, 37, 38, 39 \\
12& Swift J1753.2--0127&$ 5.3\pm0.6       $& $4-8        $& $40-55$ & 2005 & Swift/XRT & $0.5-10$ & 40, 41, 42 43\\
  &                    &                   &              &        &      & RXTE/PCA & $2-25$ & \\
13& Swift J1842.5+1124 &$ 8^*           $& $5^*      $& $45^*$ & 2008 & Swift/XRT & $0.5-10$ & 44 \\
  &                    &                   &              &        &      & RXTE/PCA & $3-25$ & \\
14& XTE J1118+480      &$ 7.0\pm0.7       $& $1.8        $& $68-79$ & 2000, 2005 & RXTE/PCA & $3-25$ & 45, 46, 47, 48  \\
15& XTE J1550--564     &$ 10.4\pm2.3      $& $4.4\pm0.5  $& $57-77$ & 1998, 2000 & RXTE/PCA & $2-20$ & 49, 50, 51, 52, 53\\
  &                    &                   &              &        &      & RXTE/HEXTE & $20-200$ & \\ 
16& XTE J1652--453     &$ 8^*          $& $5^*      $& $45^*$ & 2009 & Swift/XRT & $0.5-10$ & 54 \\
  &                    &                   &              &        &      & RXTE/PCA & $2-25$ & \\
17& XTE J1748--288     &$ 8^*          $& $5^*      $& $45^*$ & 1998 & RXTE/PCA & $3-25$ & 55  \\
18& XTE J1817--330     &$ 8^*          $& $5^*      $& $45^*$ & 2006 & RXTE/PCA & $2-25$ & 56, 57  \\
19& XTE J1908+094      &$ 6.5\pm0.7       $& $10\pm1     $& $27\pm3$ & 2002-03 & RXTE/PCA & $2-25$ & 59, 60, 61, 62 \\
20& XTE J2012+381      &$ 8^*             $& $5^*      $& $45^*$ & 1998 & RXTE/PCA & $3-20$ & 63  \\
\hline
\end{tabular}
\leftline{$^*$ Estimation not available, values assumed in this work.}
\leftline{Errors are quoted when available.}
\leftline{(1) \citet{KC2021}, (2) \citet{della1994}, (3) \citet{Tao2019}, (4) \citet{Bassi2019}. 
(5) \citet{Sreehari2019}, (6) \citet{Parker2016},}
\leftline{(7) \citet{Dincer2012}, (8) \citet{Nandi2012}, (9) \citet{Shui2021}, (10) \citet{Molla2017}, (11) \citet{Steiner2012}, (12) \citet{McClintock2009},}
\leftline{(13) \citet{Kalemci2006}, (14) \citet{Capitanio2009}, (15) \citet{Chen2010}, (16) \citet{Zhou2013}, (17) \citet{AJ2021a}, (18) \citet{Lazar2021},}
\leftline{(19) \citet{AJ2020a}, (20) \citet{Chauhan2021}, (21) \citet{Jia2022}, (22) \citet{Shang2019}, (23) \citet{Chauhan2019}, (24) \citet{Xu2018},} 
\leftline{(24) \citet{Tao2018}, (25) \citet{Stiele2018}, (26) \citet{Molla2016}, (27) \citet{Kuulkers2013}, (28) \citet{Torres2021}, (29) \citet{DD2015a}, }
\leftline{(30) \citet{Alabarta2020}, (31) \citet{AJ2021b}, (32) \citet{Armas2019}, (33) \citet{AJ2016}, (34) \citet{Russell2014a}, (35) \citet{Russell2014b},}
\leftline{(36) \citet{Mondal2019}, (37) \citet{Armas2014}, (38) \citet{Stiele2018}, (39) \citet{Armas2013} (40) \citet{DD2017},}
\leftline{(41) \citet{CadolleBel2007}, (42) \citet{Neustroev2014}, (43) \citet{Shaw2016}, (44) \citet{Zhao2016}, (45) \citet{DC2019}, (46) \citet{McClintock2001},}
\leftline{(47) \citet{Khargharia2013}, (48) \citet{DD2020}, (49) \citet{Orosz2002}, (50) \citet{Orosz2011}, (51) \citet{Kreidberg2012}, (52) \citet{Sobczak2000},}
\leftline{(53) \citet{Rodriguez2003}, (54) \citet{Han2011}, (55) \citet{Revnivtsev2000}, (56) \citet{Rykoff2007}, (58) \citet{Gierlinski2008}, (59) \citet{DC2021}, }
\leftline{(60) \citet{Curran2015}, (61) \citet{Draghis2021}, (62) \citet{Gogus2004}, (63) \citet{Vasiliev2000}.}
\label{tab:list}
\end{table*}

\section{Sample Selection and Data Acquisition}
\label{sec:data}

\subsection{Sample}
\label{sec:sample}
We selected the sources from the literature where the outburst evolution of the sources is studied. We chose the sources where the information of inner disc temperature ($T_{\rm in}$), disc normalization ($N_{\rm DBB}$) or inner disc radius ($R_{\rm in}$), photon index ($\Gamma$), disc flux ($F_{\rm disc}$), Comptonized flux ($F_{\rm Compt}$) and line flux ($F_{\rm line}$) are available. In total, we selected 32 outbursts of 20 sources for our study. Out of a total of 20 sources, we selected one outburst each for 16 sources, two outbursts each for two sources, and six outbursts for two sources. GX 339--4 and H 1743--322 showed $\sim 8-10$ outbursts in the last 25 years. However, we chose only six outbursts for each source where the information on spectral evolution was available. We did not include the data of any persistent sources in this work. The information about the outburst is tabulated in Table~\ref{tab:list}.

\subsection{Mass, Distance and Inclination}
\label{sec:mass}
We collected the information on the mass ($M_{\rm BH}$), distance ($d$) and inclination angle ($i$) of the sources from the literature. The information of the mass, distance and inclination angle are available from 15, 15 and 13 sources, respectively. When no information is available, we assumed $M_{\rm BH} = 8$ $M_{\odot}$, $d=5$~kpc and $i=45$ degrees in our study. The details information is presented in Table~\ref{tab:list}. 

\subsection{Spectral Parameters}
\label{sec:parameter}
In the present study, we concentrated on the spectral properties of BHXBs. In this work, we studied the disc, Comptonized and iron line emission of BHXBs. As most of the studies were done using RXTE/PCA data up to 20~keV, the study of the reflection was scarce. However, iron $K\alpha$ line, which is produced via reflection, is detected at $\sim 6.4$~keV. The spectra were analysed using a multi-colour disc blackbody (MCD), a Comptonized component and iron line emission. The MCD emission was modelled with the {\tt XSPEC} model \textsc{diskbb} \citep{Mitsuda1984,Makishima1986} and the Comptonized emission was modelled with the {\tt XSPEC} model \textsc{power-law} (PL), \textsc{cutoff power-law} (CPL) or \textsc{compTT} model \citep{T94}. The iron line emission was modelled with a Gaussian component at $\sim 6.4$~keV. 

From these spectral modelling, we accuired the information of inner disc temperature ($T_{\rm in}$), photon index ($\Gamma$), cut-off energy ($E_{\rm cut}$), disc flux ($F_{\rm disc})$, Comptonized flux ($F_{\rm Compt}$) and line flux ($F_{\rm line}$). From the \textsc{diskbb} normalization ($N_{\rm dbb}$), we obtained the inner disc radius ($R_{\rm in}$). The $N_{\rm dbb}$ is given by, $N_{\rm dbb} = (r_{\rm in}/d_{\rm 10})^2 \cos{i}$, where $r_{\rm in}$, $d_{\rm 10}$ and $i$ are the apparent inner disc radius, distance in 10~kpc and disc inclination angle. The apparent radius ($r_{\rm in}$) is related to the true inner disc radius ($R_{\rm in}$) as $R_{\rm in} = \xi \kappa^2 r_{\rm in}$, where $\xi$ and $\kappa$ are correction factor and spectral colour correction factor, respectively \citep[e.g.,][]{Shimura-Takahara1995,Kubota1998}. We used $\xi=0.41$ \citep{Kubota1998} and $\kappa=1.8$ for our calculation.

The $E_{\rm cut}$ is related to the $kT_{\rm e}$ as $E_{\rm cut} \sim 2-3~kT_{\rm e}$ \citep[e.g.,][]{Petrucci2001}. In our analysis, we collected the information of $kT_{\rm e}$ and $E_{\rm cut}$ in the HCS from the literature. We used $E_{\rm cut}=2kT_{\rm e}$ in our analysis, which is a reasonable approximation. The reflection fraction often measures the strength of the reflection ($R_{\rm refl}$), which is defined as the ratio of the reflected emission to the direct emission to the observer. In this study, we used the ratio of the iron line flux to the Comptonized flux ($F_{\rm line}/F_{\rm Compt}$) as a proxy of the reflection fraction.

\subsection{Luminosity and Bolometric Correction}
\label{sec:bol-corr}
We collected the information on the disc flux ($F_{\rm disc}$) and Comptonized flux ($F_{\rm Compt}$) from the literature. In most cases, the fluxes were reported in a narrow energy band of $0.5-20$~keV or similar. One needs to study the spectra in a broad energy range to gain insightful information. Hence, we calculated the bolometric flux in the energy range of $0.1-500$~keV by applying a bolometric correction to the fluxes obtained in a limited energy band. \citet{Maccarone2003} applied bolometric correction assuming the spectra in the form of $\frac{dN}{dE} \sim E^{-1.8} \exp^{-E/200~{\rm keV}}$ in the energy range of $0.5~{\rm keV}-10~{\rm MeV}$. They noted that if $\Gamma$ is changed by $0.1$, the flux will change by $10\%$, which is less than other uncertainties. \citet{Vahdat2019} calculated disc flux in the range of $0.01-200$~keV and power-law flux in the range of $T_{\rm in}$ to $200$~keV. \citet{Dunn2010} calculated the disc flux in the energy range of $0.001-100$~keV and power-law flux in the $1-100$~keV. 

Here, we calculated the disc flux in the $0.001-10$~keV and Comptonized power-law flux in the range between $T_{\rm in}$ to $500$~keV, when the disc was detected. When the disc was not detected, we calculated the Comptonized flux in the energy range of $0.1-500$~keV. As the disc is unlikely to contribute significantly over $\sim 5-6$~keV, we only calculated the disc luminosity up to 10~keV. To calculate the Comptonized flux, we applied bolometric correction assuming the spectra in the form of $\frac{dN}{dE} \sim E^{-\Gamma} \exp^{-E/200~{\rm keV}}$. Once we calculated the bolometric disc and Comptonized flux, we calculated the luminosity using, $L=4\pi d^2 F$, where $d$ is the source distance. The total luminosity is calculated as $L_{\rm tot} = L_{\rm disc}^{0.001-10~{\rm keV}} + L_{\rm Compt}^{0.1-500~{\rm keV}}$. The hardness ratio (HR), disc fraction ($f_{\rm disc}$) and Comptonized fraction ($f_{\rm Compt}$) were calculated as HR$= L_{\rm Compt}/L_{\rm disc}$, $f_{\rm disc}=L_{\rm disc}/L_{\rm tot}$, $f_{\rm Compt}=L_{\rm Compt}/L_{\rm tot}$, respectively.

We calculated Eddington scaled luminosity or Eddington ratio ($\lambda_{\rm Edd}$) by taking the ratio of the luminosity to the Eddington luminosity ($L_{\rm Edd}$), i.e. $\lambda_{\rm Edd}=L/L_{\rm Edd}$. The Eddington ratio is given by $L_{\rm Edd}=1.3\times 10^{38}~(M_{\rm BH}/M_{\sun})$. We calculated total Eddington ratio ($\lambda_{\rm tot}$), disc Eddington ratio ($\lambda_{\rm disc}$) and Comptonized Eddington ratio ($\lambda_{\rm Compt}$) as $\lambda_{\rm tot} = L_{\rm tot}/L_{\rm Edd}$, $\lambda_{\rm disc} = L_{\rm disc}/L_{\rm Edd}$ and $\lambda_{\rm Compt} = L_{\rm Compt}/L_{\rm Edd}$, respectively.

\begin{figure*}
\centering
\includegraphics[width=8.5cm]{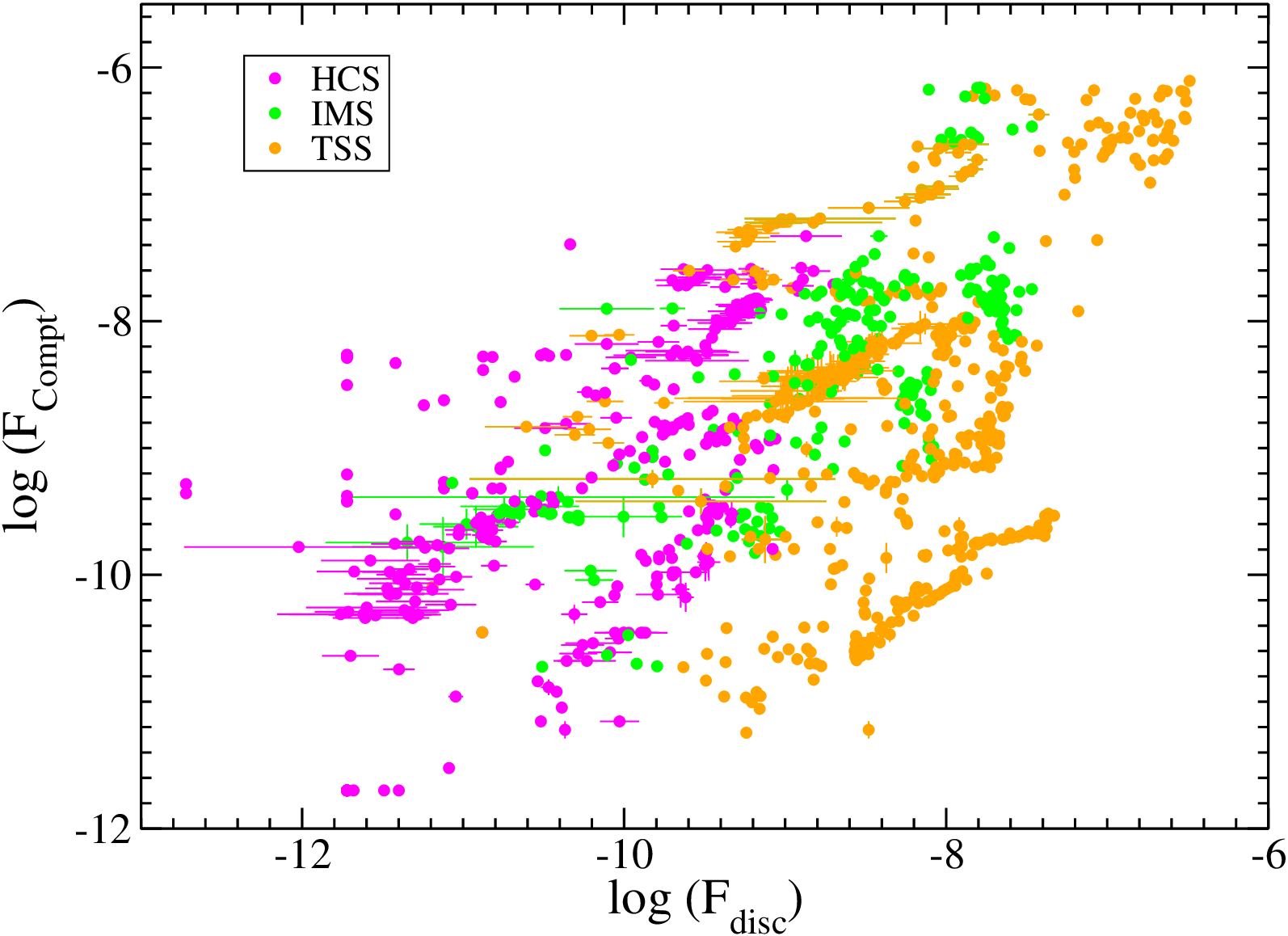}
\includegraphics[width=8.5cm]{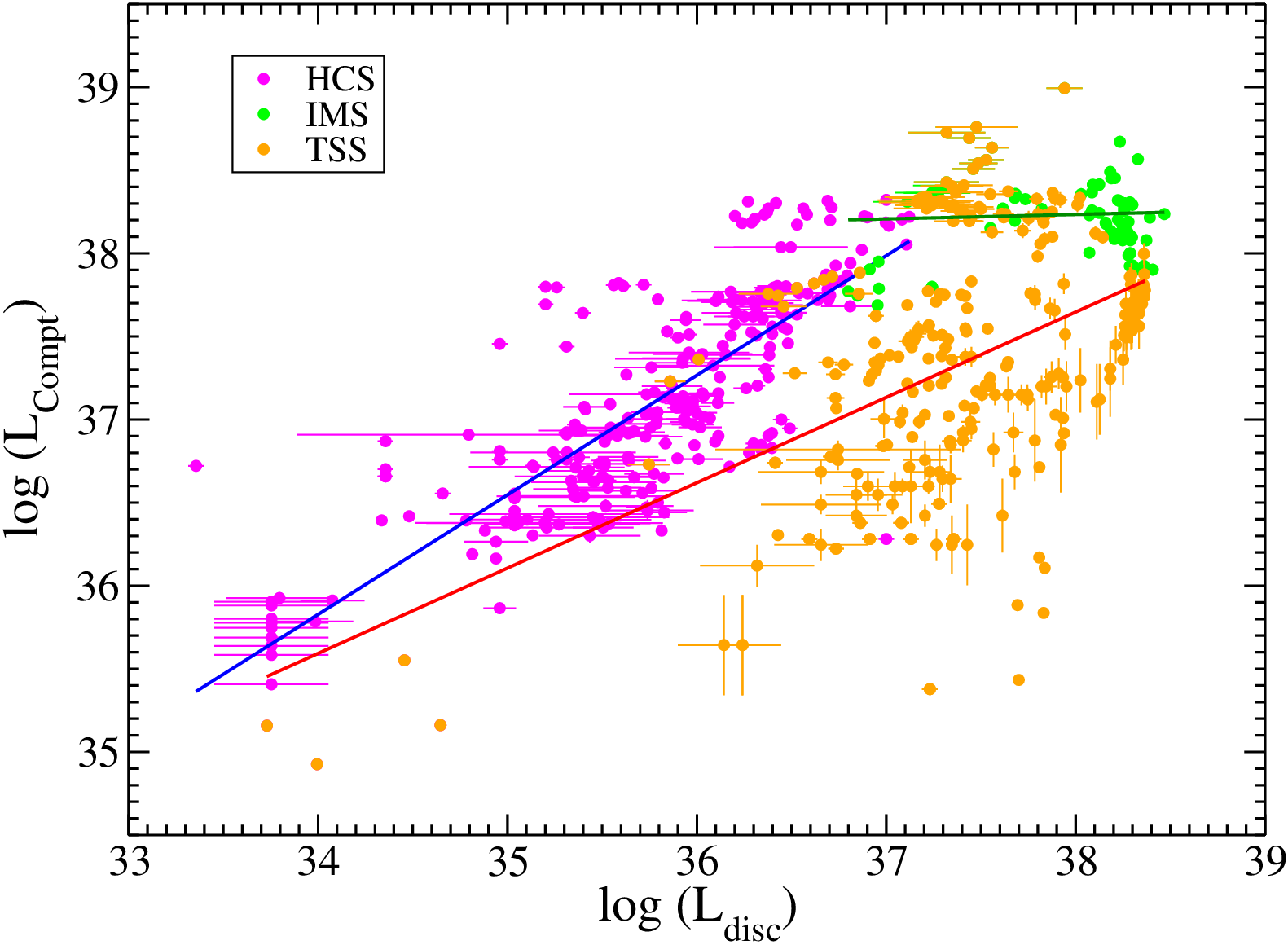}
\caption{Left panel: the Comptonized flux ($F_{\rm Compt}$) plotted as a function the disc flux ($F_{\rm disc}$). Right panel: the Comptonized luminosity ($L_{\rm Compt}$) is plotted as a function of the disc luminosity ($L_{\rm disc}$). The magenta, green and orange points represent the observations from the HCS, IMS and TSS, respectively. The blue, dark green and red solid lines represent the linear best fit for the HCS, IMS and TSS, respectively, in the right panel.}
\label{fig:disk-comp-lum-hs}
\end{figure*}

\section{Analysis and Result}
\subsection{Flux and Luminosity}
The standard thin disc emits thermal soft photons. A fraction of the thermal photons intercept in the Compton corona, and in turn, the Compton corona produces the Comptonized power-law spectra via inverse-Comptonization. Thus, in a way, the Comptonized emission is related to the thermal emission. The left panel of Figure~\ref{fig:disk-comp-lum-hs} shows the variation of the $F_{\rm Compt}$ as a function of the $F_{\rm disc}$. In the right panel of Figure~\ref{fig:disk-comp-lum-hs}, we plot the $L_{\rm Compt}$ as a function of the $L_{\rm disc}$. The magenta, green and orange points represent the data from the HCS, IMS and TSS, respectively. In the right panel of Figure~\ref{fig:disk-comp-lum-hs}, the blue, dark green and red solid lines represent the linear best fit in the HCS, IMS and TSS, respectively. We excluded the sources with unknown distances in the right panel of Figure~\ref{fig:disk-comp-lum-hs}. In our sample, out of 32 outbursts, the HCS, IMS and TSS were observed in 31, 20 and 14 outbursts, respectively. In many observations in the HCS, the disc was not detected. Those observations were excluded in Figure~\ref{fig:disk-comp-lum-hs}. 

We found that the $F_{\rm disc}$ and $F_{\rm Compt}$ are moderately correlated in all the spectral states, with the Pearson correlation coefficient $r=0.671$ with the p-value of $<10^{-4}$ in the HCS, $r=0.541$ with the p-value of $<10^{-4}$ in the IMS, and $r=0.457$ with the p-value of $<10^{-4}$ in the TSS, respectively. In terms of luminosity, we observed a strong correlation between $L_{\rm disc}$ and $L_{\rm Compt}$ in the HCS and a moderate correlation in the TSS. The Pearson correlation coefficient between $L_{\rm disc}$ and $L_{\rm Compt}$ was observed to be $r=0.805$ with the p-value of $<10^{-4}$ and $r=0.556$ with the p-value of $<10^{-4}$ in the HCS and TSS, respectively. Surprisingly, no correlation was observed in the IMS.

We also conducted a similar study using the Eddington ratio ($\lambda$), replacing the luminosity and obtained a similar result. We found that $\lambda_{\rm disc}$ and $\lambda_{\rm Compt}$ are strongly correlated in the HCS with the Pearson correlation coefficient as $r=0.794$ with the p-value of $<10^{-4}$. In the TSS, a moderate correlation was found with the Pearson correlation coefficient of $r=0.589$ with the p-value of $<10^{-4}$. As before, no correlation was observed in the IMS.

\begin{figure*}
\centering
\includegraphics[width=8.5cm]{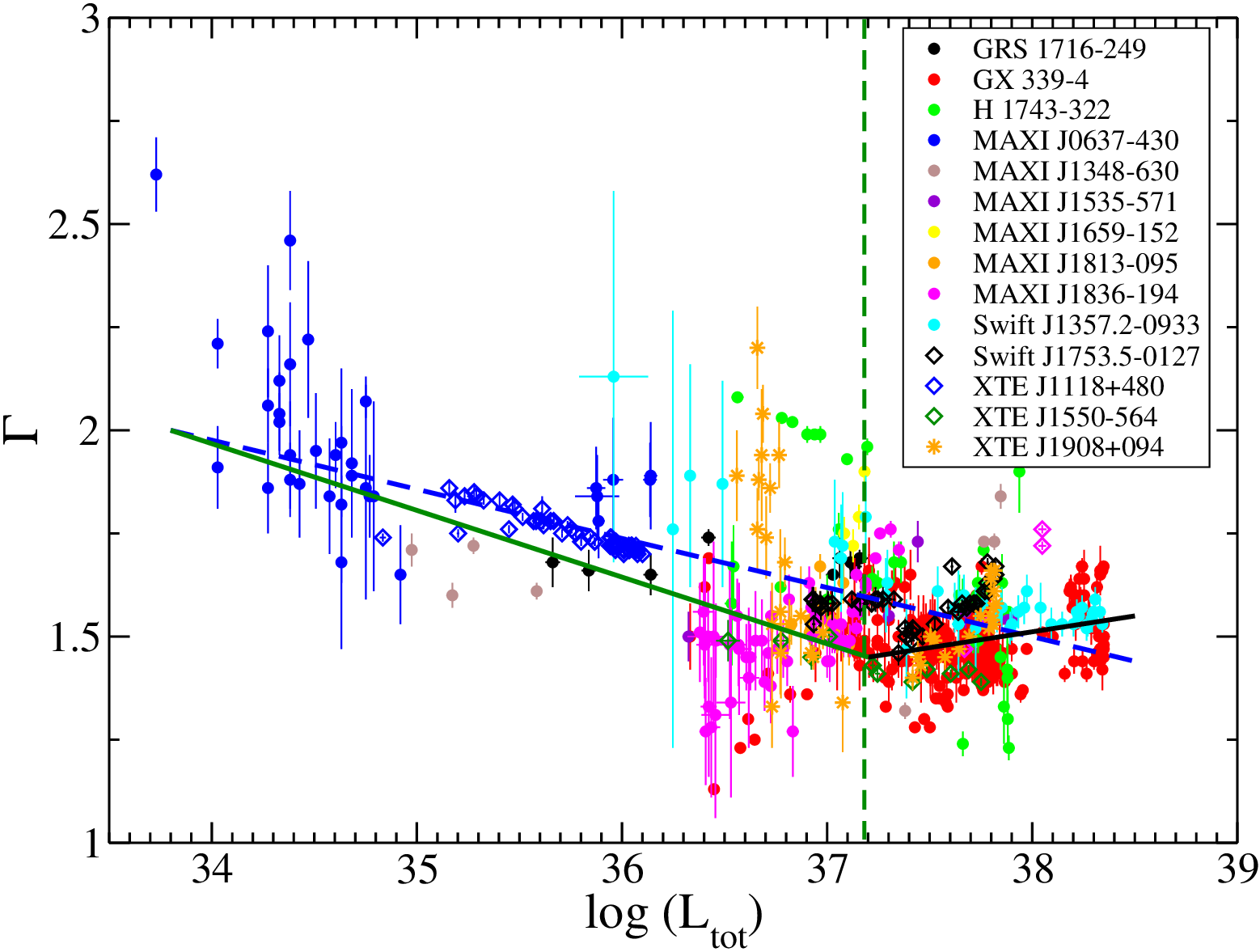}
\includegraphics[width=8.5cm]{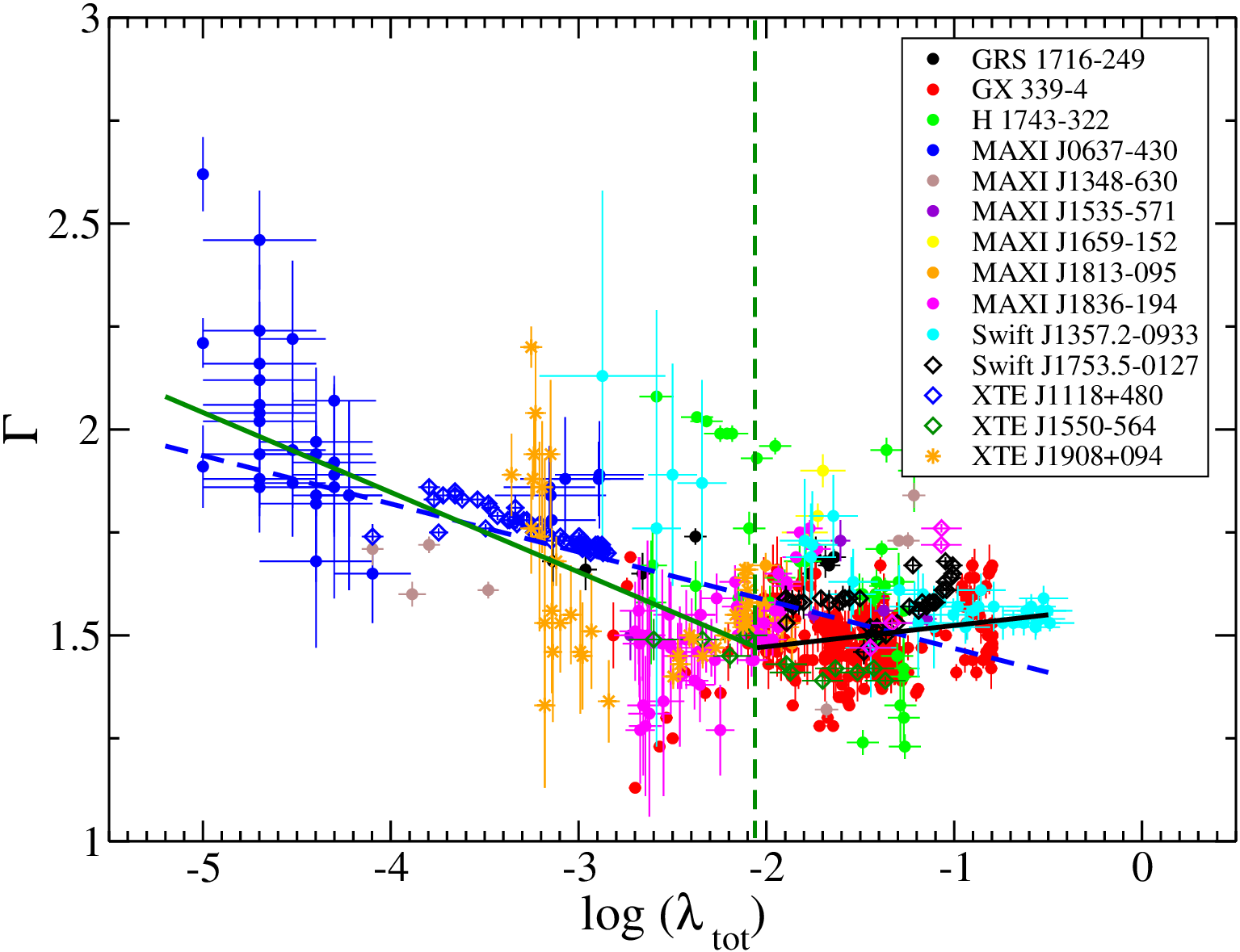}
\caption{Left panel: the photon index ($\Gamma$) is plotted as a function of the total luminosity ($L_{\rm tot}$) in the HCS. The blue dashed line represents the linear best fit. The dark green and black lines represents the best-fit line for $L_{\rm tot}<L_{\rm t}$ and $L_{\rm tot}>L_{\rm t}$, respectively. The dark green vertical line represents the transition luminosity ($L_{\rm t}$). The sources with known distances are only considered in this plot. Right panel: the variation of the photon index ($\Gamma$) is shown as a function of the total Eddington ratio ($\lambda_{\rm tot}$) for the HCS. The blue dashed line represents the linear best fit. The dark green and black lines represent the best-fit line for $\lambda_{\rm tot}<\lambda_{\rm t}$ and $\lambda_{\rm tot}>\lambda_{\rm t}$, respectively. The dark green vertical line represents the transition Eddington ratio ($\lambda_{\rm t}$). Different colours and symbols correspond to the different sources mentioned in the inset of each figure. The sources with known masses and distances are considered in this plot.}
\label{fig:lhs-gam}
\end{figure*}

\subsection{Photon Index}
The HCS data were available for 31 outbursts from 19 sources in our sample. The thermal disc emission was not detected in $\sim 60$\% observations within the sample. When the disc was detected, the disc fraction ($L_{\rm disc}/L_{\rm tot}$) was always $<25$\%. We found that the $\Gamma$ is anti-correlated with the $L_{\rm Compt}$ and the $L_{\rm tot}$ with the Pearson correlation coefficient $r=-0.599$ with the p-value of $<10^{-4}$ and $r=-0.608$ with the p-value of $<10^{-4}$, respectively, in the HCS. We did not observe any correlations of the $\Gamma$ with $L_{\rm Compt}$ and $L_{\rm tot}$ in the TSS and IMS. We also checked if the $\Gamma$ is correlated with the HR or Comptonized fraction ($f_{\rm Compt}$) across the spectral states. We did not find any correlation/anti-correlation between $\Gamma$ and HR or $f_{\rm Compt}$ in any spectral states. A similar result was found while studying the correlation of $\Gamma$ with $\lambda_{\rm Compt}$ and $\lambda_{\rm tot}$.

The left panel of Figure~\ref{fig:lhs-gam} shows the variation of the $\Gamma$ as a function of the $L_{\rm tot}$ in our sample. We did not include the sources with unknown distances in the plot. The blue dashed line represents the linear fit of the data. From the linear regression, we obtained $\Gamma = -(0.12\pm0.006) L_{\rm tot}+(6.05\pm0.24)$. The Pearson correlation coefficient between $\Gamma$ and $L_{\rm tot}$ was found to be $r = 0.518$ with the p-value of $<10^{-4}$. This indicated a moderate negative correlation between these two parameters. We noticed that the anti-correlation weakens at the higher luminosity, indicating a `V'-shaped correlation. Thus, we fitted the data in two regions of luminosity. We used the following linear relation for regression analysis,

\begin{equation}
\label{eqn:lin}
\begin{array}{cc}
Y(X)=A_1 (X-X_t)+B_1~~~(X<X_t) \\
~~~~~~~~~~=A_2 (X-X_t)+B_2~~~(X>X_t).
\end{array}
\end{equation}

Here, replacing Y(X), X and X$_t$ with $\Gamma$, $\log L_{\rm tot}$ and $\log (L_t)$, we obtained a negative correlation for the low-luminosity region ($L_{\rm tot} < L_t$) and a positive correlation for the high luminosity region ($L_{\rm tot} > L_t$). The slope was found to flip at $\log L_{\rm t} \approx 37.2$, i.e. at $L_{\rm t} \approx 1.5\times 10^{37}$ \eps. We obtained the slope of the fits as $A_1=-0.16\pm0.02$ for $L \lesssim 10^{37.2}$ and $A_2=0.03\pm0.01$ for $L_{\rm tot} \gtrsim 10^{37.2}$. In the left panel of Figure~\ref{fig:lhs-gam}, the vertical line represents the transition luminosity ($L_{\rm t}$). The dark green and black lines represent the best-fit line for $L_{\rm tot}<L_{\rm t}$ and $L_{\rm tot}>L_{\rm t}$, respectively.

The right panel of Figure~\ref{fig:lhs-gam} shows the variation of the $\Gamma$ as a function of the $\lambda_{\rm tot}$. We did not include sources with unknown masses and distances in this plot. Using {\rm linear regression analysis}, we obtained a negative correlation between the $\Gamma$ and $\lambda_{\rm tot}$. The linear regression analysis returned as $\Gamma = (-0.12\pm0.006)\lambda_{\rm tot} +(1.35\pm0.10)$. We obtained the Pearson correlation coefficient as $r=-0.608$ with the p-value $<10^{-4}$. Similar to $\Gamma-L_{\rm tot}$ relation, the correlation of $\Gamma-\lambda_{\rm tot}$ is also observed to flip. Replacing Y(X), X and X$_t$ with $\Gamma$, $\log \lambda_{\rm tot}$ and $\lambda_t$ in Equation~\ref{eqn:lin}, we observed that the turnover happens at $\log \lambda_{\rm t} \approx -2$. We obtained the slope as $A_1 = 0.19\pm0.02$ for $\lambda_{\rm tot}<\lambda_t$ and $A_2 = 0.04\pm0.02$ for $\lambda_{\rm tot}>\lambda_t$. In the right panel Figure~\ref{fig:lhs-gam}, the blue dashed line represents the linear fit of the data. The vertical line represents the transition Eddington ratio ($\lambda_{\rm t}$). The dark green and black lines represent the best-fit line for $\lambda_{\rm tot}<\lambda_{\rm t}$ and $\lambda_{\rm tot}>\lambda_{\rm t}$, respectively.

We repeated the above exercise for $L_{\rm Compt}$ and $\Gamma$. The $\Gamma$ was observed to be anti-correlated with the $L_{\rm Compt}$ with the Pearson correlation coefficient $r=-0.599$ with the p-value of $<10^{-4}$ in the HCS. Linear regression analysis returned as $\Gamma = (-0.12\pm0.006)+(6.00\pm0.23)$. Similar to the $\Gamma-L_{\rm tot}$ correlation, two different correlation branches were observed in $\Gamma-L_{\rm Compt}$ relation. The turnover was observed at $L_{\rm t, Compt} \simeq 10^{37.15}$, i.e., $L_{\rm t, Compt} \simeq 1.4\times 10^{37}$ \eps. The linear slope was obtained as $-0.15\pm0.01$ for $L_{\rm Compt}<L_{\rm t, Compt}$ and $0.003\pm0.021$ for $L_{\rm Compt}>L_{\rm t, Compt}$. No correlation between $\Gamma$ and $L_{\rm Compt}$ was found in the IMS and TSS.

\begin{figure*}
\centering
\includegraphics[width=8.5cm]{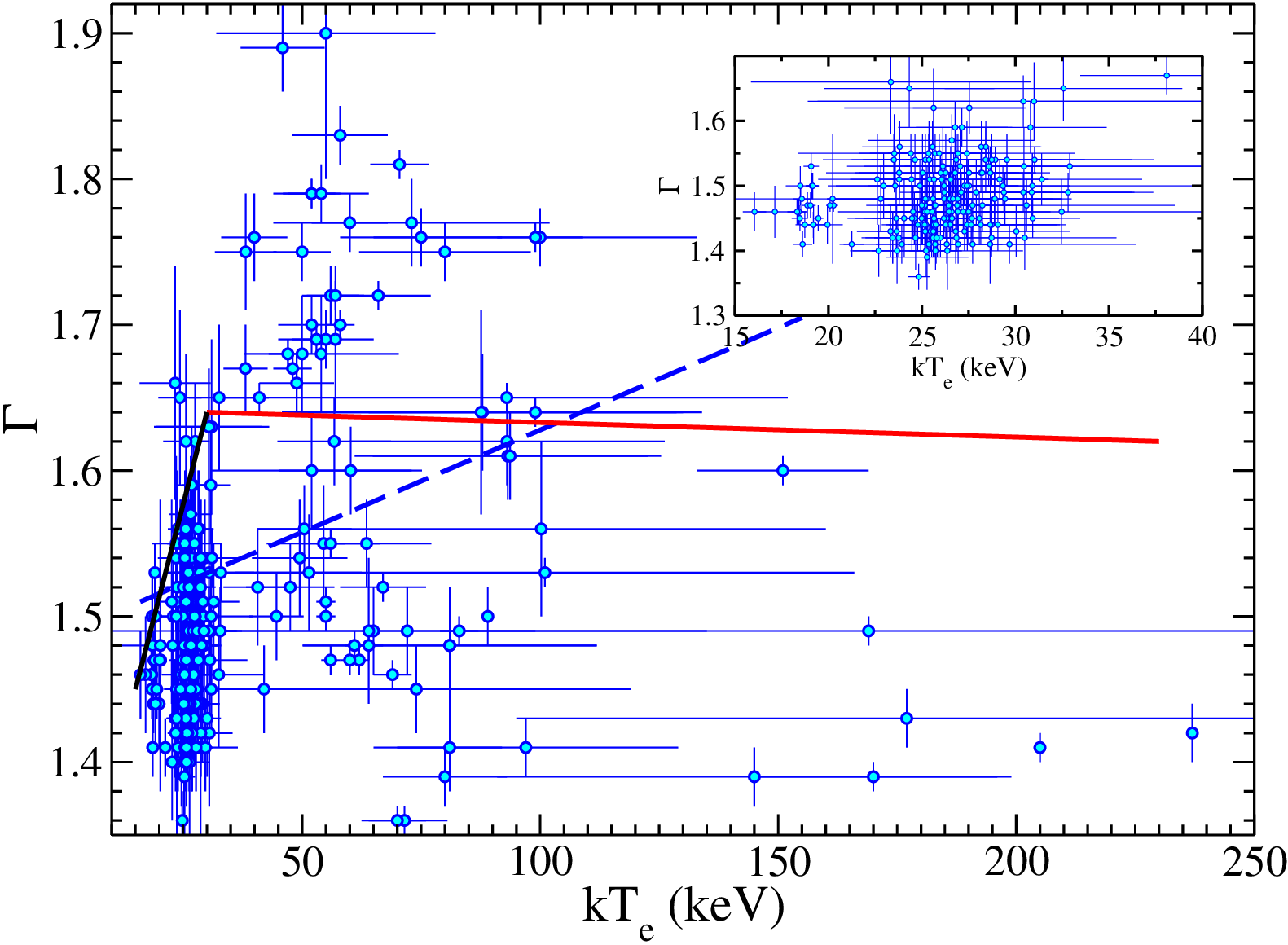}
\includegraphics[width=8.5cm]{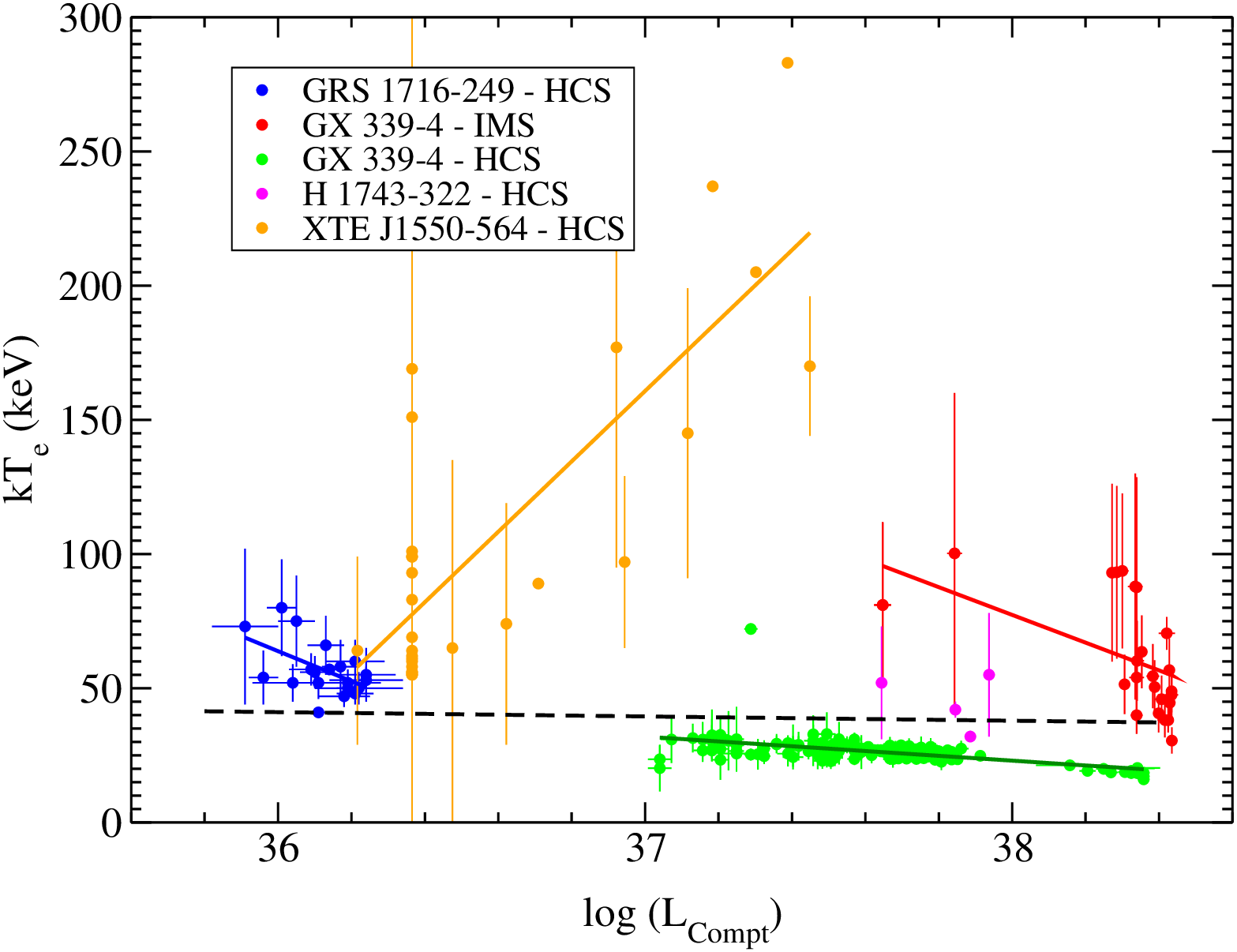}
\caption{Left panel: variation of the photon index ($\Gamma$) is plotted as a function of the hot electrons temperature ($kT_{\rm e}$). The blue dashed, solid black and solid red lines represent the linear best-fit for the entire data, for $kT_{\rm e}<30$~keV and for $kT_{\rm e}>30$~keV, respectively. The inset figure show the zoomed-in for clarity. Right panel: the variation hot electrons temperature ($kT_{\rm e}$) is shown as a function of the Comptonized luminosity ($L_{\rm Compt}$). The black dashed, solid blue, solid red, solid dark green, solid magenta and solid orange line represent the best linear fit for the entire sample, GRS 1716--249 in the HCS, GX 339--4 in the HCS, GX 339--4 in the IMS, and XTE J155--564 in the HCS, respectively. The blue, red, green, magenta and orange circles represent the data points for GRS 1716--249 in the HCS, GX 339--4 in the HCS, GX 339--4 in the IMS, H 1743--322 in the HCS, and XTE J155--564 in the HCS, respectively.}
\label{fig:kt}
\end{figure*}

\subsection{Hot Electron's Temperature}
The information of $kT_{\rm e}$ or $E_{\rm cut}$ were available for four BHXBs in our sample. Figure~\ref{fig:kt} shows the variation of the $\Gamma$ with the $kT_{\rm e}$ in the left panel. The Pearson correlation coefficient between $\Gamma$ and $kT_{\rm e}$ is found to be $0.25$ with the p-value $<10^{-4}$, indicating no correlation. A linear fitting gives $\Gamma = (0.001\pm0.003) kT_{\rm e} + (1.49\pm0.16)$. It seems that two different correlations exist for the high and low-temperature regions. We performed similar linear regression analysis as in \S3.2, to check the different correlations of $\Gamma$ and $kT_{\rm e}$ in two different regions of $kT_{\rm e}$. We obtained the slope of linear relation as $A_{1}=0.002\pm0.001$ for $kT_{\rm e}<30$~keV and $A_{2}=-0.002\pm0.006$ for $kT_{\rm e}>30$~keV, respectively. The turn-over of correlation in $\Gamma-kT_{\rm e}$ relation was not statistically significant. In the left panel of Figure~\ref{fig:kt}, the blue dashed, solid black and solid red lines represent the linear best-fit for the entire data, for $kT_{\rm e}<30$~keV and  for $kT_{\rm e}>30$~keV, respectively.

In the right panel of Figure~\ref{fig:kt}, we plot $kT_{\rm e}$ as a function of the $L_{\rm Compt}$. We did not find any correlation between $kT_{\rm e}$ and $L_{\rm Compt}$ in our sample. The Pearson correlation coefficient was $r=0.25$ with a p-value of $0.403$. A linear regression analysis returned with $\log L_{\rm Compt}=(-1.61\pm0.32) kT_{\rm e} + (99.2 \pm 19)$, indicating a negative correlation. For individual sources, the negative correlation was stronger than the entire sample. In the right panel of Figure~\ref{fig:kt}, the blue, green, red, magenta and orange circles represent the data of GSR 1716--249 in the HCS, GX 339--4 in the HCS, GX 339--4 during the state transition/IMS, H1743--322 in the HCS and XTE J1550--564 in the HCS, respectively. Except for XTE J1550--564, a negative correlation between the $kT_{\rm e}$ and $L_{\rm Compt}$ was observed for the rest of the sources. We obtained the Pearson correlation coefficient $r=-0.53$ with the p-value of $0.01$ for GRS 1716--249, $-0.540$ with p-value of $0.004$ for GX 339--4 in the IMS, $-0.587$ with the p-value of $<0.001$ for GX 339--4 in the HCS, $0.797$ with the p-value of $<0.001$ for XTE J1550--564, respectively. We did not calculate the correlation coefficient for H1743--322 as fewer than 11 data points were available. A linear regression analysis returned with different slopes for different sources, with $-56.1\pm21.0$ for GRS 1716--249 in the HCS, $51.7\pm 16.4$ GX 339--4-in the IMS, $-9.0\pm 0.9$ for GX 339--4 in the HCS, and $131.2 \pm 19.5$ for XTE J1550+564 in the HCS. As different source show different correlations, no correlation between the $kT_{\rm e}$ and $L_{\rm Compt}$ is seen. We also studied the correlation of $kT_{\rm e}$ with the $\lambda_{\rm Compt}$. We obtained a similar correlation of $kT_{\rm e}$ with $\lambda_{\rm Compt}$ as $L_{\rm Compt}$.

\begin{figure*}
\centering
\includegraphics[width=16cm]{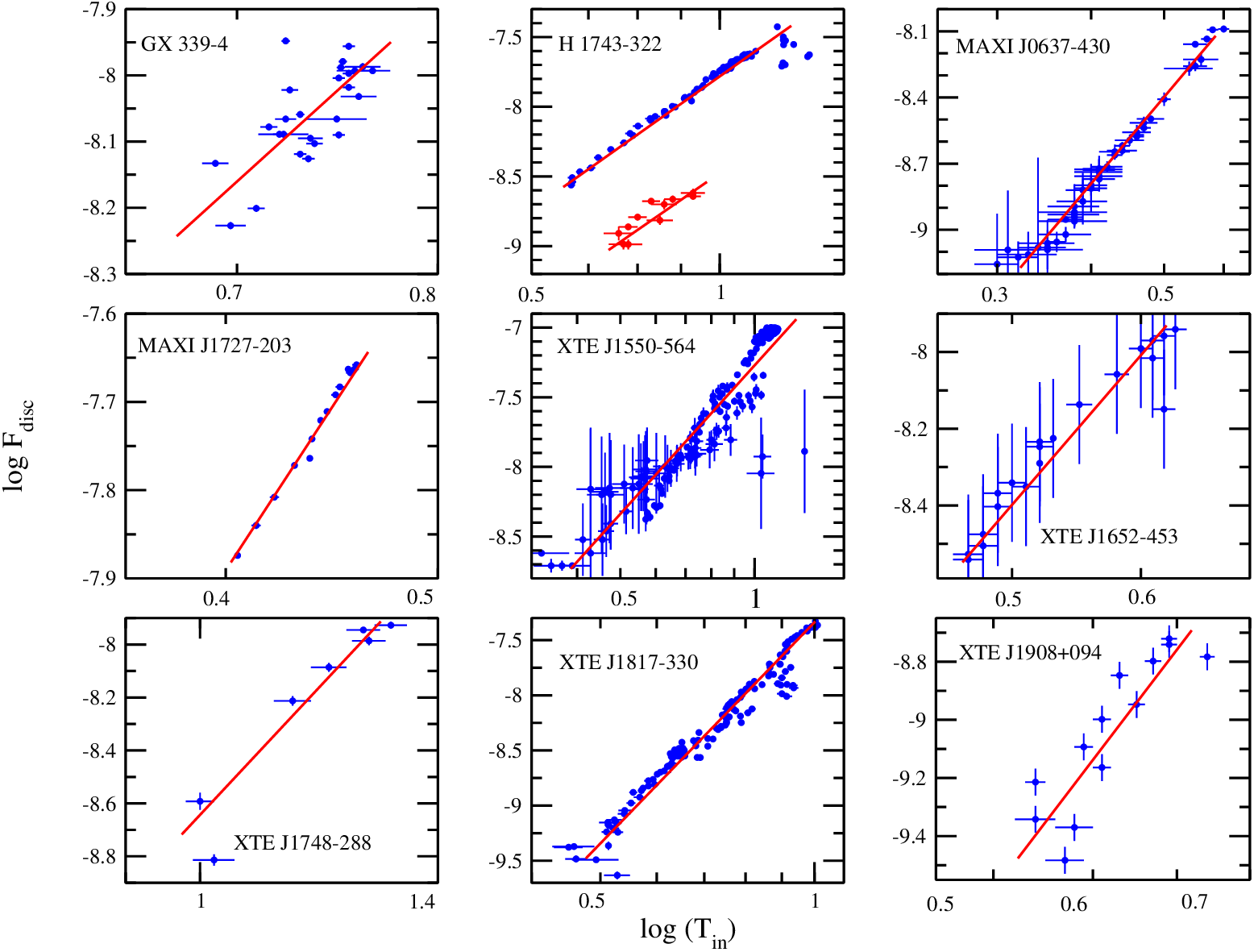}
\caption{The disc flux ($F_{\rm disc}$) is plotted as a function of the inner disc temperature ($T_{\rm in}$) for nine BHXBs. Solid red lines represent the best-fit by liner regression using $\log F_{\rm disc}=A*\log T_{\rm in}+B$.}
\label{fig:tin}
\end{figure*}

\subsection{Disc Temperature}
The disc flux is expected to vary with the $T_{\rm in}$ as $F_{\rm disc} \propto T_{\rm in}^4$. This relation holds if the $R_{\rm in}$ and the colour correction factor ($\kappa$) remain constant. We expect this relation to hold in the TSS as the disc extends to the innermost stable circular orbit (ISCO). Figure~\ref{fig:tin} shows the variation of the $F_{\rm disc}$ as a function of the $T_{\rm in}$ for nine sources in the TSS. In each panel, the solid red line represents the best fit by the linear regression method. The fit was done using the relation, $\log(F_{\rm disc})=A*\log(T_{\rm in})+B$. The details result of the linear fit is presented in Table~\ref{tab:tin}.

Out of a total of 20 sources, nine sources show TSS during their respective outbursts. Of these nine sources, $F_{\rm disc} \propto T_{\rm in}^4$ hold for the six sources in the TSS. Three sources, namely, XTE J1748--288, XTE J1817--330 and XTE J1908+094 showed deviation from the standard $F_{\rm disc} \propto T_{\rm in}^4$ relation.

\begin{table}
\centering
\caption{$F_{\rm disc}-T_{\rm in}$ Relation}
\begin{tabular}{ccc}
\hline
Source & A & B \\
\hline
GX 339--4 &  $4.27\pm0.85$ & $-7.48\pm0.18$\\
H 1743--322  & $3.78\pm0.12$ & $-7.81\pm0.13$ \\
MAXI J0637--430 & $4.00\pm0.10$ & $-7.20\pm0.14$ \\
MAXI J1727--203 & $4.03\pm0.10$ & $-6.29\pm0.16$\\
XTE J1550--564  & $3.54\pm0.10$ & $-7.28\pm0.31$\\
XTE J1652--223  & $4.29\pm0.29$ & $-7.07\pm0.15$\\
XTE J1748--288  & $6.14\pm0.94$ & $-8.71\pm0.32$\\
XTE J1817--330  & $6.07\pm0.10$ & $-7.42\pm0.14$\\
XTE J1908+094   & $5.31\pm0.89$ & $-7.94\pm0.19$\\
\hline
\end{tabular}
\leftline{The result of the linear regression of the $L_{\rm disc}$ and $T_{\rm in}$. The data are fitted }
\leftline{with $\log(F_{\rm disc})=A*\log(T_{\rm in})+B$ in the TSS.}
\label{tab:tin}
\end{table}

\begin{figure}
\centering
\includegraphics[width=8.5cm]{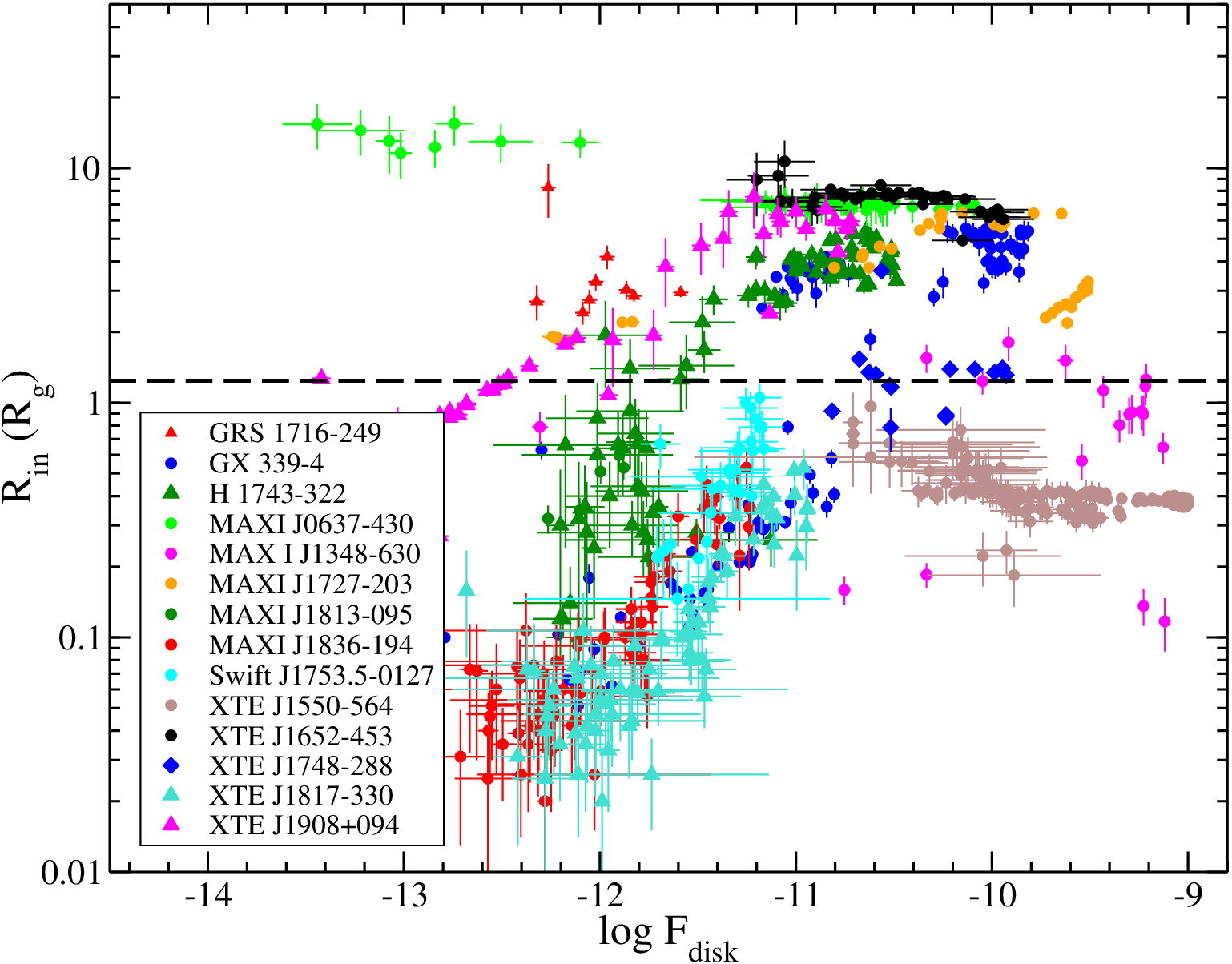}
\caption{Inner disc radius ($R_{\rm in }$) plotted as a function of the disc flux ($F_{\rm disc}$). Different points represent different sources.}
\label{fig:rin}
\end{figure}

\begin{figure*}
\centering
\includegraphics[width=8.5cm]{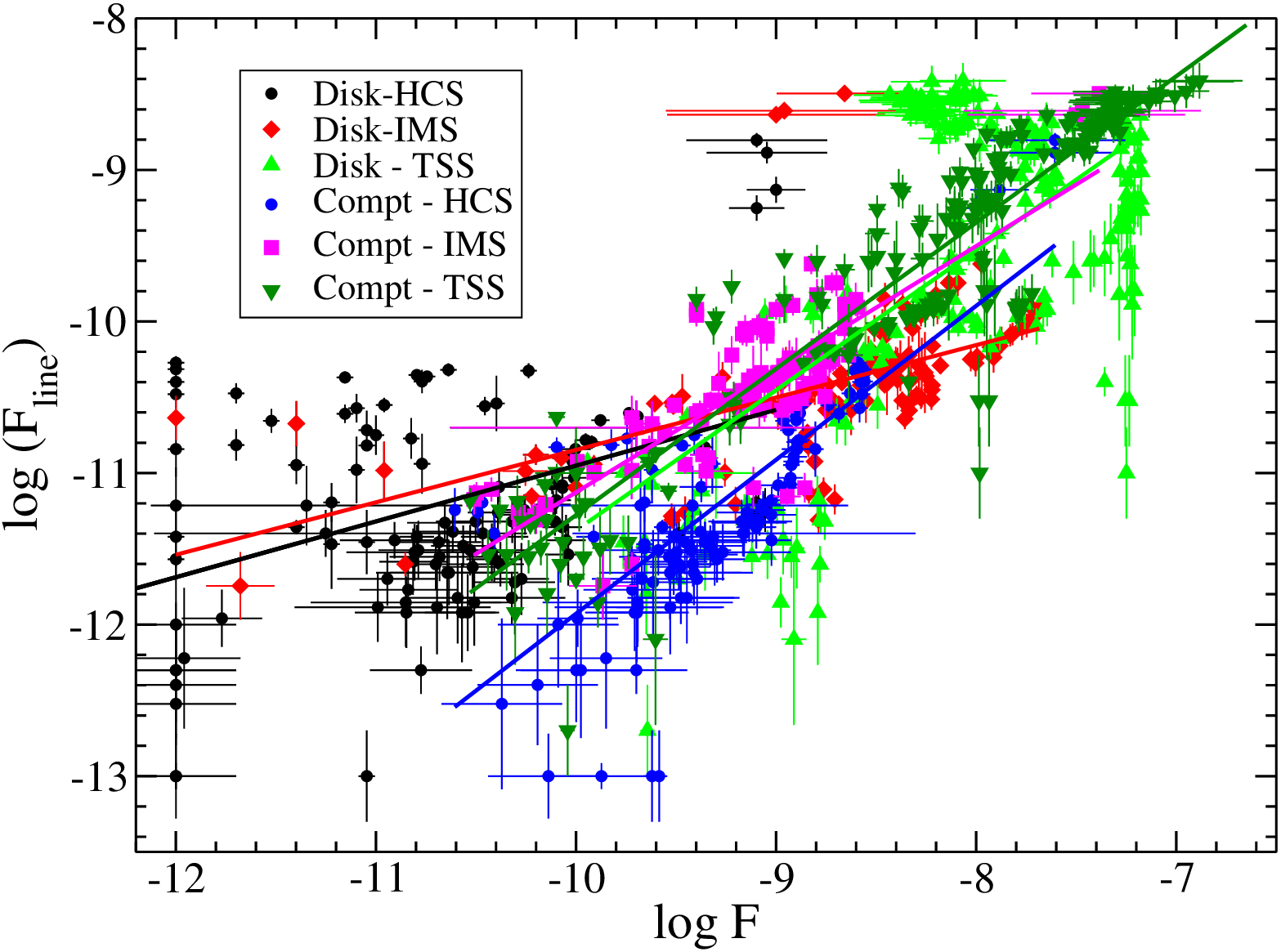}
\includegraphics[width=8.5cm]{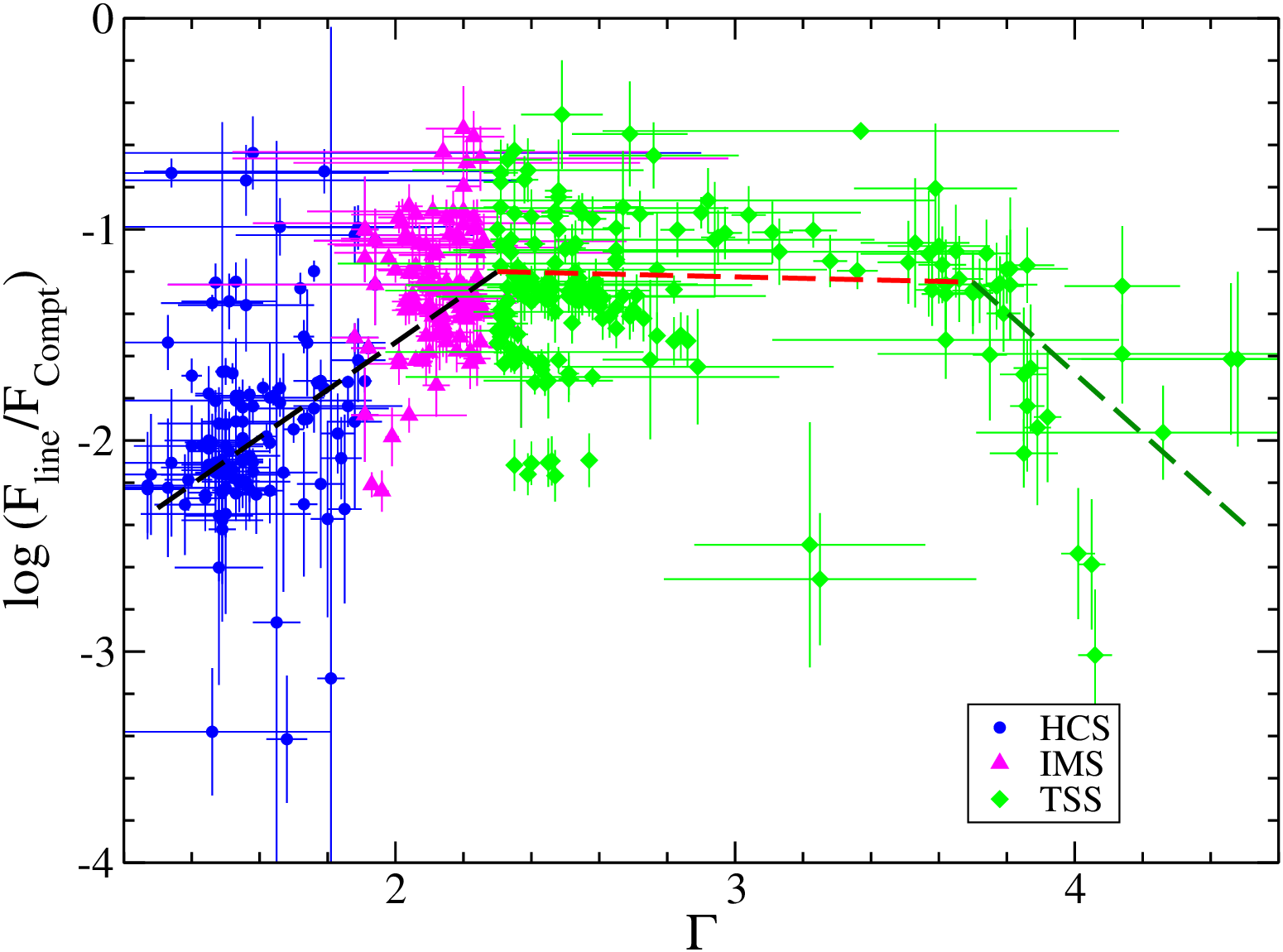}
\caption{Left panel: variation of the line flux ($F_{\rm line}$) is plotted as a function of disc ($F_{\rm disc}$) and Comptonized flux ($F_{\rm Compt}$) in the HCS, IMS and TSS. The black circles, red diamonds and green up-triangles represent the disc flux in the HCS, IMS and TSS, respectively. The Comptonized flux in the HCS, IMS and TSS are shown by blue circles, magenta squares and dark green down-triangles, respectively. The solid black, red and green lines represent the linear best-fit for $F_{\rm disc}$ in the HCS, IMS and TSS, respectively, while the solid blue, magenta and dark green lines represent the linear best-fit for the $F_{\rm Compt}$ in the HCS, IMS and TSS, respectively. Right panel: the variation of the ratio of the line flux to the Comptonized flux ($F_{\rm line}/F_{\rm Compt}$) is shown as a function of the photon index ($\Gamma$), in the HCS, IMS, and TSS, respectively. The blue circles, magenta triangles and green circles represent the data from the HCS, IMS and TSS, respectively. The black dashed, red dashed and dark green dashed lines represent the linear best fit for the region $\Gamma<2.3$, $2.3< \Gamma <3.9$, and $\Gamma>3.9$, respectively.}
\label{fig:fe}
\end{figure*}

\subsection{Inner disc Radius}
The $R_{\rm in}$ can be obtained from the \textsc{diskbb} normalization (see Section~\ref{sec:parameter}), assuming a Schwarzschild black hole. During an outburst, an evolving disc is often observed with the disc moving in or out. Here we studied the variation of the $R_{\rm in}$ with the $F_{\rm disc}$. In our sample, the information of the $R_{\rm in}$ was available for 14 sources. We show the variation of $R_{\rm in }$ with the $F_{\rm disc}$ in Figure~\ref{fig:rin}. The horizontal black dashed line represents $R_{\rm in}=1.24~R_{\rm G}$, which is the location of ISCO for a maximally rotating BH. Many points lie below the $R_{\rm in}=1.24~R_{\rm G}$ line, which is unphysical. For $R_{\rm in}>1.24~R_{\rm G}$, we did not observe any clear patterns of the variation of $R_{\rm in}$ with $F_{\rm disc}$. Although some sources, e.g., MAXI J0637--430, XTE J1652--253, showed a negative correlation while others, e.g., MAXI J1727--203, XTE J1908+094, indicated a positive correlation.

\subsection{Iron Line \& Reprocessed Emission}
The left panel of Figure~\ref{fig:fe} shows the variation of the $F_{\rm line}$ as a function of the $F_{\rm disc}$ and $F_{\rm Compt}$ in the HCS, IMS and TSS. The black circles, red diamonds and green up-triangles denote the disc flux in the HCS, IMS and TSS, respectively. The Comptonized flux in the HCS, IMS and TSS are shown by blue circles, magenta squares and dark green down-triangles, respectively. The solid black, red and green lines represent the linear best-fit for $F_{\rm disc}$ in the HCS, IMS and TSS, respectively, while the solid blue, magenta and dark green lines represent the linear best-fit for the $F_{\rm Compt}$ in the HCS, IMS and TSS, respectively. In the right panel of Figure~\ref{fig:fe}, We show the variation of $F_{\rm line}/F_{\rm Compt}$ as a function of the $\Gamma$.  The blue, magenta and green circles represent the data points from the HCS, IMS and TSS, respectively. The black, red and green dashed lines represent the linear best fit for $\Gamma<2.3$, $2.3<\Gamma<3.9$, and $\Gamma>3.9$, respectively.

In our sample, the iron line was detected for 10 BHXBs. We found that the $F_{\rm line}$ is correlated with the $F_{\rm disc}$ and $F_{\rm Compt}$ for the whole sample with the Pearson correlation coefficient of $0.738$ with the p-value of $<10^{-4}$ and $0.897$ with the p-value of $<10^{-4}$, respectively. However, when we considered the data within individual spectral states, the correlation of $F_{\rm disc}$ and $F_{\rm line}$ breaks down slightly. The $F_{\rm disc}$ was found to be moderately correlated with the $F_{\rm line}$; with the Pearson correlation coefficient: $r=0.418$ with the p-value of $<10^{-4}$ in the HCS, $r=0.554$ with the p-value of $<10^{-4}$ in the IMS, and $r=0.634$ with the p-value of $<10^{-4}$ in the TSS. The $F_{\rm Compt}$ was observed to be strongly correlated with the $F_{\rm line}$ in all spectral states; with the Pearson correlation coefficient $r=0.768$ with the p-value of $<10^{-4}$ in the HCS, $r=0.854$ with the p-value of $<10^{-4}$ in the IMS, and $r=0.926$ with the p-value of $<10^{-4}$ in the TSS.

We also conducted linear regression analysis of $F_{\rm line}$ with the $F_{\rm disc}$ and $F_{\rm Compt}$. A linear regression analysis yields the slope of $1.03\pm 0.05$ and $0.71\pm0.02$ for $F_{\rm diskc}$ and $F_{\rm Compt}$, respectively, for the entire sample. Considering the individual spectral state, we obtained the slope as $0.37\pm0.09$ in the HCS, $0.35\pm0.09$ in the IMS and $0.92\pm0.18$ in the TSS for $F_{\rm disc}$. For $F_{\rm Compt}$, the slopes are obtained as $1.02\pm0.07$ in the HCS, $0.81\pm0.05$ in the IMS and $0.97\pm0.03$ in the TSS.

We found that the $\Gamma$ is moderately correlated with the $F_{\rm line}/F_{\rm Compt}$ with $r=0.346$ with the p-value of $<10^{-4}$ in the HCS. No correlation was found between the $\Gamma$ and $F_{\rm line}/F_{\rm Compt}$ in the IMS and TSS, respectively. When we performed linear regression analysis, we found the slope as $1.10\pm0.08$ for $\Gamma<2.3$, $-0.01\pm0.05$ for $2.3<\Gamma<3.9$ and $1.73\pm0.91$ for $\Gamma>3.9$. 

\section{Discussion and Concluding Remarks}
We collected spectral data from 32 outbursts of 20 BHXBs from the literature. We checked for correlations among different spectral parameters to understand the accretion evolution across different spectral states on a global scale.

A BHXB is generally observed in the HCS at the beginning of the outburst. As the accretion rate increases, the source evolves through various spectral states. Both $F_{\rm disc}$ and $F_{\rm Compt}$ evolve during an outburst. Generally, at the start of the outburst, $F_{\rm disc}$ is low, increasing as the source evolves and becoming maximum in the TSS. The $F_{\rm disc}$ depends on the $R_{\rm in}$ and the $T_{\rm in}$ as $F_{\rm disc} \propto R_{\rm in}^2 T_{\rm in}^4$ \citep[e.g.,][]{SS73,Dunn2011}. In the TSS, the $F_{\rm disc}$ varies as $F_{\rm disc} \propto T_{\rm in}^4$, given that $R_{\rm in}$ and the colour correction factor ($\kappa$) remains constant \citep[eg.,][]{Shimura-Takahara1995}. In our sample, the data for the TSS were available for nine sources, and out of these nine sources, six sources closely followed this relation. Moving disc or changing $\kappa$ could be the reason for not following the theoretical relation for the rest three sources. In the IMS and HCS, no sources were found to follow this relation. This is expected as the disc may move in the IMS and HCS \citep[e.g.,][]{Gierlinski2004,Dunn2010,Dunn2011}.

The evolution of the $R_{\rm in}$ is still debated. It is generally accepted that the $R_{\rm in}$ extends to the ISCO in the TSS; however, it is unclear in the HCS and IMS. Our study did not observe any clear pattern of the variation of $R_{\rm in}$ with the $F_{\rm disc}$. For some sources, e.g., MAXI J0637--430, XTE J1652--253, $R_{\rm in}$ was found to correlate negatively with the $F_{\rm disc}$, i.e., the disc moved outward when the outburst started. For some other sources, e.g., MAXI~J1727--203, XTE~J1908+094, $R_{\rm in}$ was observed to positively correlate with the $F_{\rm disc}$, i.e., the disc moved closer to the BH during the outburst. However, in our sample, the variation of the $R_{\rm in}$ is unclear for most sources. Hence, the current study could not confirm the evolution of the inner disc.

It is well known that the $T_{\rm in}$ and $N_{\rm dbb}$ are degenerate. In many observations, $T_{\rm in}$ was higher in the HCS with very small $N_{\rm dbb}$. These are not physical. Most of the studies were done using {\it RXTE}/PCA, which had a limited band-pass at the lower end of the spectra, which might cause the inaccurate estimation of the $T_{\rm in}$ and $N_{\rm DBB}$ or $R_{\rm in}$, particularly in the HCS. In general, the disc emits the maximum amount of power at $\sim 3~kT_{\rm in}$. The \swift/XRT or \nicer extends up to $\sim 0.5$ keV while {\it RXTE}/PCA extends $\sim 2$ keV at the lower end of the spectra. PCA only detects a small part of the disc emission, much beyond the Wien peak in most cases (especially in the HCS), and thus, contrary to the XRT, is not appropriate to estimate the disc properties precisely. Additionally, \textsc{diskbb} considers a non-rotating BH. Thus, non relativistic model such as \textsc{diskbb} may not calculate the $R_{\rm in}$ accurately. Hence, robust relativistic modelling in a broad energy band is required to study the evolution of the disc accurately.

A fraction of the seed photons intercepted with the corona and produced the hard Comptonized photons via inverse-Comptonization \citep[e.g.,][]{ST80,ST85,Z96}. The Comptonized spectra is generally characterized by the spectral slope or photon index ($\Gamma$). In the HCS and TSS, we observed that the $L_{\rm Compt}$ is strongly correlated with the $L_{\rm disc}$. This suggested that as the disc luminosity or the number of the seed photons increase, more photons intercept at the corona, producing higher Comptonized emission.
 
The accretion geometry is observed to differ in the HCS from the TSS. The evidence comes from the different correlation of $\Gamma$ and $L_{\rm tot}$ in the HCS and TSS \citep[e.g.,][]{Wu2008,Yang2015}. When the accretion rate is high, the standard disc supplies the seed photons to the corona, resulting in efficient cooling. This leads to a drop in the hot electron temperature and softer spectra. Thus, the thermal Comptonization naturally explains the positive correlation of $\Gamma-L_{\rm tot}$ and the negative correlation of $kT_{\rm e}-L_{\rm tot}$. However, when the accretion rate drops, the inner accretion disc is replaced by a hot accretion flow or radiatively inefficient flow \citep[e.g.,][]{Yuan2004,Yuan2014}. In this state, most of the seed photons are non-thermal and originate in the corona itself or jet \citep[e.g.,][]{Markoff2001,Yang2015}. This explains the negative correlation of $\Gamma-L_{\rm tot}$ and positive correlation of $kT_{\rm e}-L_{\rm tot}$ \citep[e.g.,][]{Yang2015,Yan2020}.

We observed a negative correlation between the $\Gamma$ and $L_{\rm Compt}$ for $L_{\rm Compt}<1.4\times 10^{37}$ \eps while a positive correlation is observed for $L_{\rm Compt}>1.4\times 10^{37}$ \eps. In terms of the total luminosity, the turn-over occurs at $L_{\rm tot} \approx 1.5\times 10^{37}$ \eps. \citet{Yan2020} found the turn-over at slightly lower luminosity at $L_{\rm tot} \sim 9 \times 10^{36}$ \eps. The turn-over of the correlation or `V'-shaped correlation of $\Gamma-L_{\rm tot}$ indicates that the disc photons still play an important role in the HCS, if not a dominant one.

We did not observe any correlation of the $\Gamma$ and $kT_{\rm e}$ in our sample. Along with the $kT_{\rm e}$, the $\Gamma$ of the emergence spectra also depends on the optical depth ($\tau$) of the Compton cloud. The non-correlation between $\Gamma$ and $kT_{\rm e}$ has been observed in other BHXBs \citep[e.g.,][]{Miyakawa2008}. Generally, it is expected that the high $kT_{\rm e}$ would result in hard spectra, i.e. $\Gamma$ would anti-correlated with the $kT_{\rm e}$, only if $\tau$ remains constant.

We also did not find any correlation between the $kT_{\rm e}$ and $L_{\rm Compt}$ in our sample. This is largely because one source behaves differently. Actually, different correlations are found for each source. A negative correlation between the $kT_{\rm e}$ and $L_{\rm Compt}$ was observed for GX 339--4, GRS 1716--249 and H 1743--322, while a positive correlation was observed for XTE~J1550--564. Both negative and positive correlation of $kT_{\rm e}$ and $L_{\rm Compt}$ has been observed in BHXBs \citep[e.g.,][]{Zdziarski2004,Yamaoka2005}.

We observed that the $F_{\rm line}$ is correlated with both the $F_{\rm disc}$ and  the $F_{\rm Compt}$, indicating the reprocessed emission depend on both disc and coronal emission. The correlation of the $F_{\rm line}$ is stronger in the TSS, implying that the reprocessing is stronger if the disc is prominent. We also found that the ratio of the line flux to the Comptonized flux ($F_{\rm line}/F_{\rm Compt}$) is positively correlated with the $\Gamma$ for the region $\Gamma<2.3$, no correlation for $3.8<\Gamma<2.3$, and negatively correlation for $\Gamma>3.8.$ This result implies that the strength of the reprocessed emission increases as the source moves toward the TSS from the HCS, with the increasing disc flux. However, after a certain point, the Comptonized flux decreases, and the reflection drops due to the lack of hard photons. Similar correlation of the $\Gamma$ and reflection fraction has been observed for both BHXBs \citep[e.g.,][]{Zdziarski1999} and active galactic nuclei \citep[e.g.,][]{Ezhikode2020}.

The IMS is complex in comparison to the HCS and TSS. The IMS is considered to be the state transition phase between the HCS and TSS. The accretion geometry is distinct in the HCS and TSS and is believed to change during the IMS. We observed a correlation between $L_{\rm disc}$ and $L_{\rm Compt}$ in the HCS and TSS, but no correlation is observed in the IMS. The IMS is associated with the flaring events \citep[e.g.,][]{Fender2004,RM06}, which makes the IMS complex. A part of the X-ray emission could come from the flare \citep[e.g.,][]{AJ2017}, resulting in non-correlations among the spectral parameters. In general, we did not observe any clear pattern of the spectral parameters in the IMS in our study. Historically, the HCS and TSS are studied in detail, whereas the IMS has not been studied extensively. However, recently, some black holes are studied in the state transition phase \citep[e.g.,][]{AJ2022,Kong2021A,Cuneo2020}.

BHXBs show rich phenomenology in both spectral and timing properties. Here, we collected the data from the literature to understand phenomenology on a global scale. The black hole accretion is studied for over half a decade now. We now understand reasonably well how the accretion process works around a black hole. However, we still lack some knowledge of the black hole accretion, especially during the state transition. Additionally, the evolution of the accretion disc in the HCS is still debated. The study of BHXBs with Phenomenological models gave us a good understanding of the accretion process; however, it is far from the complete picture. To understand the BHXB completely, one needs to study the BHXB with a more physical model that considers the relativistic effect. In this work, we discuss how spectral properties of BHXB from a phenomenological perspective. Furthermore, the timing properties of BHXB from the phenomenological perspective will be studied in future.

\section*{Acknowledgements}
We thank the anonymous reviewer for constructive suggestion and comments that improved the manuscript significantly. We would also like to thank Arka Chatterjee, Claudio Ricci, Debjit Chatterjee, Dipak Debnath, Gaurava K. Jaisawal, Hsiang-Kuang Chang, Prantik Nandi, Sachindra Naik, Sandip K. Chakrabarti and Santanu Mondal for helping me to understand the subject over the year.

\section*{DATA AVAILABILITY}
We have taken the data from the literature. Appropriate references are included in the text.

\bibliographystyle{mnras}
\bibliography{ref-bha}



\appendix



\bsp	
\label{lastpage}
\end{document}